\documentclass[aps,floatfix,showpacs,nofootinbib,twocolumn]{revtex4-1}
\usepackage{amsmath}
\usepackage{amsfonts}
\usepackage{amssymb}
\usepackage{graphics,epsfig}
\usepackage{graphicx}
\usepackage{subfigure}

\begin{document}
\title{Visualizing Spacetime Curvature via Gradient Flows II: An Example of the Construction of a Newtonian analogue}
\author{Majd Abdelqader }
\email[]{majd@astro.queensu.ca}
\author{Kayll Lake}
\email[]{lake@astro.queensu.ca}
\affiliation{Department of Physics, Queen's University, Kingston,
Ontario, Canada, K7L 3N6 }
\date{\today}

\begin{abstract}
This is the first in a series of papers in which the gradient flows of fundamental curvature invariants are used to formulate a visualization of curvature. We start with the construction of strict Newtonian analogues (not limits) of solutions to Einstein's equations based on the topology of the associated gradient flows. We do not start with any easy case. Rather, we start with the Curzon - Chazy solution, which, as history shows, is one of the most difficult exact solutions to Einstein's equations to interpret physically. A substantial part of our analysis is that of the Curzon - Chazy solution itself. Eventually we show that the entire field of the Curzon - Chazy solution, up to a region very ``close" to the the intrinsic singularity, strictly represents that of a Newtonian ring, as has  long  been suspected. In this regard, we consider our approach very successful. As regrades the local structure of the singularity of the Curzon - Chazy solution within a fully general relativistic analysis, however, whereas we make some advances, the full structure of this singularity remains incompletely resolved.

\bigskip

\end{abstract}


\maketitle

\section{Introduction}
Perhaps the most familiar example of an exact solution to Einstein's equations, generated by way of a Newtonian ``analogue", is the Curzon - Chazy solution wherein the Laplacian for a point mass in a fictitious Euclidean 3 - space is used to generate an exact static axially symmetric vacuum solution of Einstein's equations. Unfortunately, the Curzon - Chazy solution bears little resemblance to a point mass. Indeed, its singularity structure appears, at present, to be at least as complicated as that of the Kerr solution.

In this paper, following the introduction given by one of us \cite{lake1}, we consider the gradient fields of the two non-differential invariants of the Curzon - Chazy solution. These alone reveal previously unknown properties of the solution. Further, following the procedure given in \cite{lake1} for the construction of strict Newtonian analogues, based on gradient flows, we suggest, and explain in a quantitative way, that a pure Newtonian ring is ``almost" a complete analogue of the Curzon - Chazy solution. Whereas our main aim here is the construction of the analogue, we have had to do a fair amount of analysis of the Curzon - Chazy solution itself.

\section{The Curzon-Chazy Metric}
The line element for a static and axially symmetric spacetime can be written in Weyl's canonical coordinates \cite{Weyl1919}
\begin{equation}
ds^2=-e^{2U}\,dt^2+e^{-2U}\left( e^{2\gamma} \left( d\rho^2+dz^2\right)+\rho^2 d\phi^2 \right) \; ,
\label{weylmetric}
\end{equation}
where $U$ and $\gamma$ are functions of $\rho$ and $z$. It is well known that these coordinates do not behave like typical cylindrical coordinates. Einstein's field equations in vacuum with zero cosmological constant give the following \textit{linear} partial differential equation for $U$,
\begin{equation}\label{U}
    \frac{\partial^2 U}{\partial \rho^2}+\frac{1}{\rho}\frac{\partial U}{\partial \rho}+\frac{\partial^2 U}{\partial z^2}=0.
\end{equation}
Equation (\ref{U}) is Laplace's equation in a Euclidean 3-space in cylindrical polar coordinates. The general solution to (\ref{U}) can be written out and the associated Einstein equations
\begin{equation}\label{gamma1}
\frac{\partial \gamma}{\partial \rho} = \rho \left( \left(\frac{\partial U}{\partial \rho}\right)^{2}+  \left(\frac{\partial U}{\partial z}\right)^{2}  \right),\;\;\;\frac{\partial \gamma}{\partial z}=  2\rho \frac{\partial U}{\partial \rho} \frac{\partial U}{\partial z}
\end{equation}
can be considered solved. Since in the weak field we have $g_{tt} \sim -(1+2U)$, it is tempting to consider a solution-generating procedure wherein one takes a known Newtonian potential $U$ (in an unphysical Euclidean 3-space), and solves (\ref{gamma1}) for $\gamma$. We then have an exact solution to Einstein's equations. Unfortunately, the resultant Weyl solution bears little similarity to the Newtonian solution and the physical meaning of most Weyl solutions so produced remain unclear. The culprit is the \textit{non-linearity} of Einstein's equations that enters via (\ref{gamma1}).

The Curzon \cite{Curzon} - Chazy \cite{Chazy} solution (CC hereafter) is one of the simplest special cases of the Weyl metric (\ref{weylmetric}).\footnote{For a review see \cite{gp}.} The potential is taken to be the Newtonian potential of a point mass ($m$) at the center of a fictitious Euclidean 3-space, $\rho = z = 0$,
\begin{equation}\label{ccu}
    U=-\frac{m}{\sqrt{\rho^2+z^2}},\;\;\;\;m>0.
\end{equation}
With (\ref{ccu}) it follows from (\ref{gamma1}) that
\begin{equation}\label{ccgamma}
    \gamma=-\frac{m^2 \rho^2}{2 (\rho^2+z^2)^2}.
\end{equation}
The resultant metric components are well defined except at $(\rho,z)=(0,0)$. The circumference of a trajectory of constant $t$, $\rho$ and $z$ is equal to $2\pi \rho \,e^{m/\sqrt{\rho^2+z^2}}$, which behaves like Euclidean cylindrical coordinates for $\rho/m$ or $ z/m \gg 1$, but the circumference diverges in the plane $z=0$ as $\rho$ goes to zero. Further, the meaning of $m$ in the CC solution is no longer obvious.

Since the CC solution is a vacuum solution, the Ricci and mixed invariants vanish. There are then only 4 Weyl invariants to consider \cite{lake1}. Moreover, since the CC solution is static, it is a purely electric spacetime (all magnetic components of the Weyl tensor vanish). We are left with only two invariants to consider,
\begin{equation}\label{w1re}
    w1R=\frac{1}{16}E_{\alpha \beta}E^{\alpha \beta}
\end{equation}
and
\begin{equation}\label{w2re}
    w2R=-\frac{1}{32}E^{\alpha}_{\beta}E^{\gamma}_{\alpha}E^{\beta}_{\gamma}.
\end{equation}

For convenience, define
\begin{equation}\label{R}
    r^2=\rho^2+z^2.
\end{equation}
We find
\begin{widetext}
\begin{equation}\label{wir}
    w1R = 2\,{m}^{2}  {\exp({{\frac {{2m}{\rho}^{2}}{{r}^{4}}}}})\frac{ \left( 3\,{r}^{6}-6\,m{r}^{5}+3\,{m}^{2}{r}^{4}+3\,{m}^{2}{\rho}^{
2}{r}^{2}-3\,{\rho}^{2}{m}^{3}r+{\rho}^{2}{m}^{4} \right)}{ {r}^{12}
 \left( {e^{{\frac {m}{r}}}} \right) ^{4}}
\end{equation}
and
\begin{equation}\label{w2r}
    w2R = 3\,{m}^{3}  {\exp({{\frac {3{m}^{2}{\rho}^{2}}{{r}^{4}}}}})
 \frac{\left( m-r \right)  \left( 2\,{r}^{6}-4\,m{r}^{5}+2\,{m}^{2}{r}^
{4}+3\,{m}^{2}{r}^{2}{\rho}^{2}-3\,{\rho}^{2}{m}^{3}r+{\rho}^{2}{m}^{4
} \right)}{ {r}^{16} \left( {e^{{\frac {m}{r}}}} \right) ^{6}}.
\end{equation}
\end{widetext}
At first glance, it would appear that $r=0$ is singular. However, along $\rho=0$, we note that
\begin{equation}\label{rhozero}
   \chi \equiv -\frac{w2R}{6}\bigg|_{\rho=0}=\left(\frac{w1R}{6}\right)^{3/2}\bigg|_{\rho=0}=\frac{{m}^{3} \left(z-m \right) ^{3}}{{z}^{12} \left( {e^{{\frac {m}{z}
}}} \right) ^{6}}
\end{equation}
and, in particular,
\begin{equation}\label{chilim}
    \lim_{z \rightarrow 0}\chi = 0.
\end{equation}
Now, whereas $\chi$ has a local minimum ($=0$) at $z=\pm m$, and a local maximum at $z=\pm(1\pm1/\sqrt{3})m$, it is clear that there is no scalar polynomial singularity along $\rho=0$. Yet, more generally, except perhaps for selected trajectories, $w1R$ and $w2R$ both diverge at $\rho=z=0$. The directional divergence of $w1R$ \footnote{The older literature refers to the Kretschmann scalar, which, as explained in \cite{lake1}, is $8w1R$ here.} was, as far as we know, first noticed by Gautreau and Anderson \cite{ga}. This observation generated much further consideration. In terms of ``polar" coordinates $(r,\theta)$ ($\rho=r\sin(\theta), z=r\cos(\theta)$), Stachel \cite{Stachel1968} showed that the area of surfaces of constant $t$ and $r$ decreases with decreasing $r$ up to a minimum (which can be shown to be $r/m \simeq 0.5389$) and then diverges as $r/m \rightarrow 0$. These surfaces can be shown to be topologically spherical. Cooperstock and Junevicus \cite{Cooperstock} showed that even trajectories of the simple form $z=C \rho^n$, where $C$ and $n$ are positive constants, give $w1R$ a rich structure.  Further analysis of the geodesics followed in order to explore the singularity and the global structure of the metric \cite{Szekeres1973}, \cite{Scott1986}, \cite{Scott1986a}, \cite{Felice1991}.\footnote{It is remarkable that the simple coordinates used by Stachel (which we refer to as Stachel, rather than ``polar" coordinates) have not been exploited further in the study of the CC metric. This is examined in Appendix A where the relevant results of Cooperstock and Junevicus, and Scott and Szekeres are generalized.} The general consensus is that the singularity has a ``ring-like", rather than ``point-like", structure. This is not so simple as it first sounds. The ``ring" has finite radius but infinite circumference \cite{Scott1986}. Rather remarkably late was the computation of $w2R$ in \cite{Arianhod1}, a work which gave visual information on the CC metric based on the principal null directions. This procedure gives much less information than the visualization procedure considered here. More recently, Taylor \cite{taylor} has suggested a technique for unravelling directional singularities.\footnote{Taylor's technique is not applicable to the CC solution due to a critical point along $\rho=0$ as $z \rightarrow 0$.}

\section{Gradient fields}

As in \cite{lake1} we define the gradient fields
\begin{equation}\label{gradient}
    k^{\alpha}_{n} \equiv - \nabla ^{\alpha}\mathcal{I}_{n}=- g^{\alpha \beta}\frac{\partial \mathcal{I}_{n}}{\partial x^{\beta}},\;\;\;k_{n\;\alpha}=-\frac{\partial \mathcal{I}_{n}}{\partial x^{\alpha}},
\end{equation}
where $n$ labels the invariant and now
\begin{equation}\label{invars}
    \mathcal{I}_{1} = w1R,\;\;\;\;\mathcal{I}_{2} = w2R.
\end{equation}

\subsection{Stachel Coordinates}
The explicit forms for the gradient fields are given in Appendix B in Stachel coordinates. Here we prefer to draw the flows explicitly in Weyl coordinates. There is no loss of information in doing this since the inverse transformations, $r=\sqrt{\rho^2+z^2}, \theta=\arctan(\rho/z)$, are so simple.\footnote{That is, to view the flow in Stachel coordinates simply think of $r$ as a circle centered on an origin at $\rho=z=0$, and $\theta$ a straight line through the origin measured from $0$ along the vertical to $\pi/2$ in the equatorial plane.}
\subsection{Weyl Coordinates\footnote{Since the flows are only two-dimensional, to find the winding number (i.e. index) of a critical point numerically, we evaluate the following integral
$$
I=\frac{1}{2\pi}\int_{C}{d\theta}
$$
where $\theta = tan^{-1} \left( \frac{v_y}{v_x} \right)$, for any vector field $\vec{v}$, and the loop $C$ can be any loop that encloses only the critical point around which the winding number is calculated.}}
\subsubsection{$\mathcal{I}_{1}$}
\begin{figure}[ht]
\epsfig{file=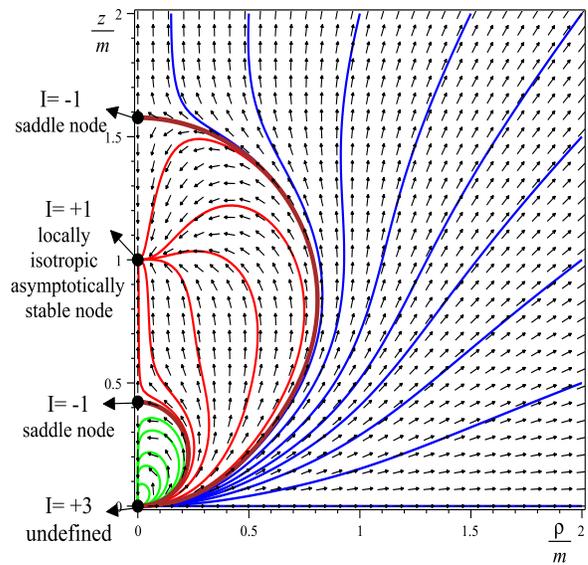,height=3in,width=4in,angle=0}
\caption{\label{curzonA}The gradient field $k_{1\;\alpha}$ of the Weyl invariant $w1R$ for the CC metric, presented in Weyl coordinates. The field is normalized for visual representation, and the flowlines are colored (online) to categorize them into distinct groups according to their global behavior (blue, red and green). Critical points are represented by black circles, and critical directions of the fields are brown. The character of the critical points along with their associated indices are shown and discussed in the text.}
\end{figure}

Along $\rho=0$ we find
\begin{equation}\label{k1w}
    k_{1\;\alpha} = \frac{12m^2(3z^3 \epsilon -9z^2m+8m^2z \epsilon-2m^3) \epsilon}{z^{10} \exp(\frac{4m \epsilon}{z})} \delta_{\alpha}^{z}
\end{equation}
where $\epsilon \equiv sign(z)$. The flow (\ref{k1w}) has 7 critical points $(\rho,z)$: $(0,0)$ (along $\rho =0 , z \rightarrow 0$), $(0,\pm m)$, $\left(0,\pm m (1-1/\sqrt{3})\right)$ and $\left(0,\pm m (1+1/\sqrt{3})\right)$. Note that the gradient field is undefined at $(0,0)$ (along $z =0 , \rho \rightarrow 0$).  To classify these critical points we calculate the Hessian
\begin{equation}\label{hessian}
    H_{\alpha \beta} \equiv -\nabla_{\alpha}k_{\beta}.
\end{equation}
Let $H$ be the determinant of $H_{\alpha \beta}$. In the full spacetime it can be shown that $H=0$ at all critical points. Similarly, in the 3 - dimensional subspace $\phi=constant$ again $H=0$. In the $\rho - z$ plane we calculate
\begin{equation}\label{hrz}
    H=-\frac{m^4h1h2}{z^{24} \exp{(\frac{8m \epsilon}{z}})}
\end{equation}
where
\begin{equation}\label{h1}
    h1 \equiv 9\,\epsilon\,{z}^{4}-36\,m{z}^{3}+42\,{m}^{2}\epsilon\,{z}^{2}-15\,{m}
^{3}z-{m}^{4}\epsilon
\end{equation}
and
\begin{equation}\label{h2}
    h2 \equiv 21\,\epsilon\,{z}^{4}-87\,m{z}^{3}+117\,{m}^{2}\epsilon\,{z}^{2}-60\,{
m}^{3}z+10\,{m}^{4}\epsilon.
\end{equation}
Since
\begin{equation}\label{w1rz}
    w1R = \frac{6\,{m}^{2} \left( {z}^{2}-2\,zm+\epsilon\,+{m}^{2}\epsilon \right)}
{\epsilon{z}^{8}  {\exp{({\frac {4m\epsilon}{z})}}}}
\end{equation}
along $\rho=0$, we are in a position to classify the critical points: $(0,\pm m)$ (asymptotically stable nodes of index $+1$ and isotropic critical points \cite{lake1} with $w1R=0$), $\left(0,\pm m (1-1/\sqrt{3})\right)$ (hyperbolic saddle points of index $-1$), and $\left(0,\pm m (1+1/\sqrt{3})\right)$ (hyperbolic saddle points of index $-1$). The overall index for any hypersurface of constant $t$ and $\phi$ is $+1$. The flow is shown in Figure \ref{curzonA} in the first quadrant only, since the metric is axially symmetric, as well as symmetric about the equatorial plane ($z \rightarrow - z$).

\subsubsection{$\mathcal{I}_{2}$}

\begin{figure}[ht]
\epsfig{file=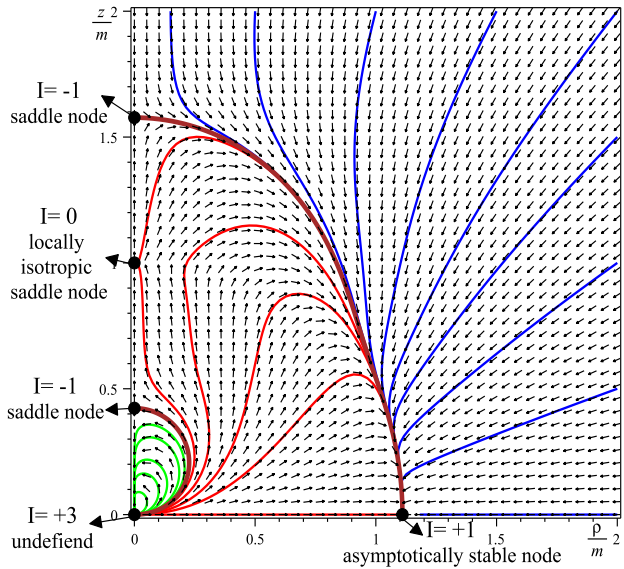,height=3in,width=4in,angle=0}
\caption{\label{curzonB}As in Figure \ref{curzonA} but for $k_{2\; \alpha}$.}
\end{figure}

Following the same procedure given above, for $\mathcal{I}_2$ along $\rho=0$ we find the critical points: $(0,\pm m)$ (degenerate isotropic critical points of index $0$), $\left(0,\pm m (1-1/\sqrt{3})\right)$ (hyperbolic saddle points of index $-1$), and $\left(0,\pm m (1+1/\sqrt{3})\right)$ (hyperbolic saddle points of index $-1$). In addition, along $z=0$ we find critical points at $\rho \cong \pm 1.1101 m$, asymptotically stable nodes of index $+1$.  The overall index for any hypersurface of constant $t$ and $\phi$ is again $+1$.\footnote{Of course the indices must be calculated in the full ``plane", for both positive and negative $z$ and for $\phi = 0$ and $\phi=\pi$.} The flow is shown in Figure \ref{curzonB}

\subsection{Scott - Szekeres Unfolding}
The unfolding of $(\rho,z)=(0,0)$, given by Scott and Szekeres in \cite{Scott1986} and \cite{Scott1986a}, is reproduced in Appendix C. Spacelike infinity ($r \rightarrow \infty$) is mapped onto $0 \leq X \leq \pi$ with $Y = \pi/2$ and $X = \pi$ with $0 \leq Y \leq \pi/2$, the $z$ axis maps onto $-\pi/2 < Y < \pi/2$ with $X=0$ and the $\rho$ axis onto $\pi/2 < X < \pi$ with $Y=0$. Now $r=0$ corresponds to $0 \leq X \leq \pi/2$ with $Y = -\pi/2$ \textit{and} $X = \pi/2$ with $-\pi/2 \leq Y \leq 0$. The singularity of the CC metric is represented only by $X = \pi/2$ with $Y = 0$.

\begin{figure}[ht]
\epsfig{file=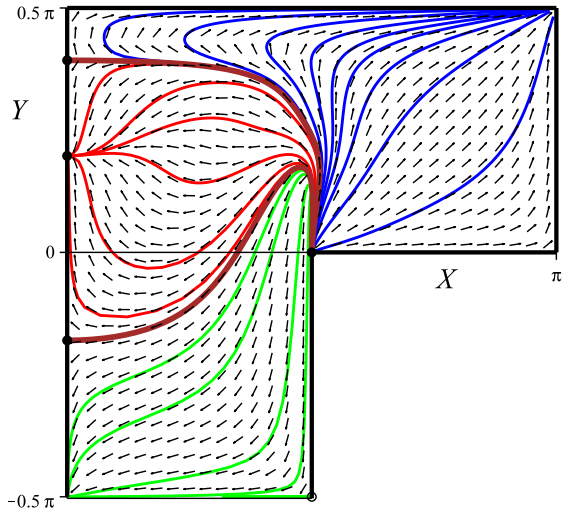,height=3in,width=3in,angle=0}
\caption{\label{curzonSSA}As in Figure \ref{curzonA} but with the Scott - Szekeres unfolding. All flow lines intersect the singularity at $(X,Y)=(\pi/2,0)$. All green flow lines also intersect $(X,Y)=(\pi/2,-\pi/2)$ which is the critical point $(\rho,z)=(0,0)$ along $\rho =0 , z \rightarrow 0$.}
\end{figure}

\begin{figure}[ht]
\epsfig{file=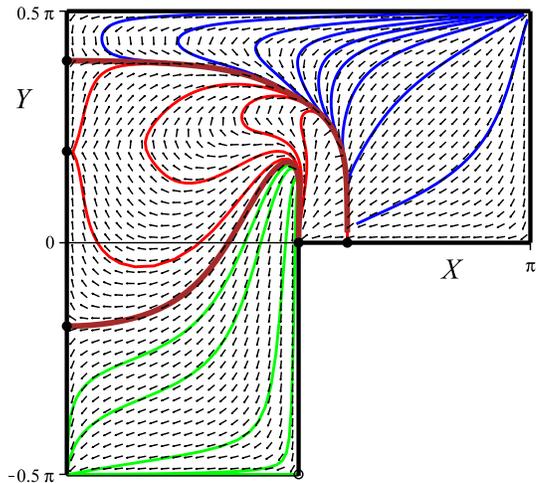,height=3in,width=3in,angle=0}
\caption{\label{curzonSSB}As in Figure \ref{curzonB} but with the Scott - Szekeres unfolding.}
\end{figure}

The gradient fields of the CC metric, with this unfolding, are shown in Figures \ref{curzonSSA} and \ref{curzonSSB}. The unfolding of the Weyl coordinate point $(0,0)$ is now evident: All of the flowlines intersect the singularity $(X,Y)=(\pi/2,0)$. The flowlines of the inner (green) region also intersect $(X,Y)=(\pi/2,-\pi/2)$.

\subsection{A New Unfolding}
\begin{figure}[ht]
\epsfig{file=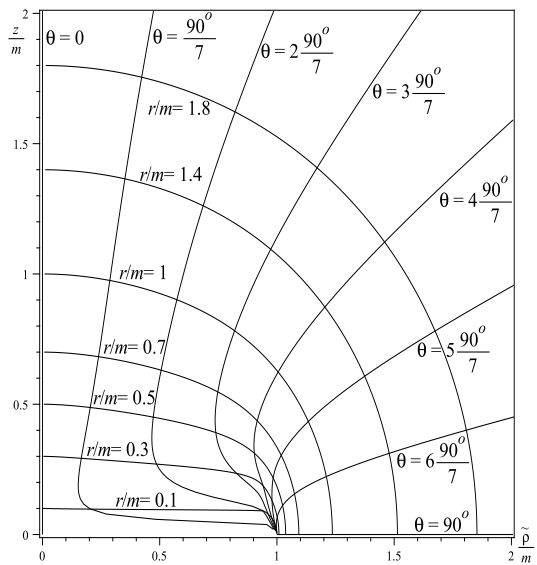,height=3in,width=3in,angle=0}
\caption{\label{newcoords}The $z/m -\tilde{ \rho}/m$ quarter plane. These are \textit{not} oblate spheroidal coordinates. The ``edge" $0 < \tilde{ \rho} < m$, where $r=0$, is not part of the spacetime.}
\end{figure}

\begin{figure}[ht]
\epsfig{file=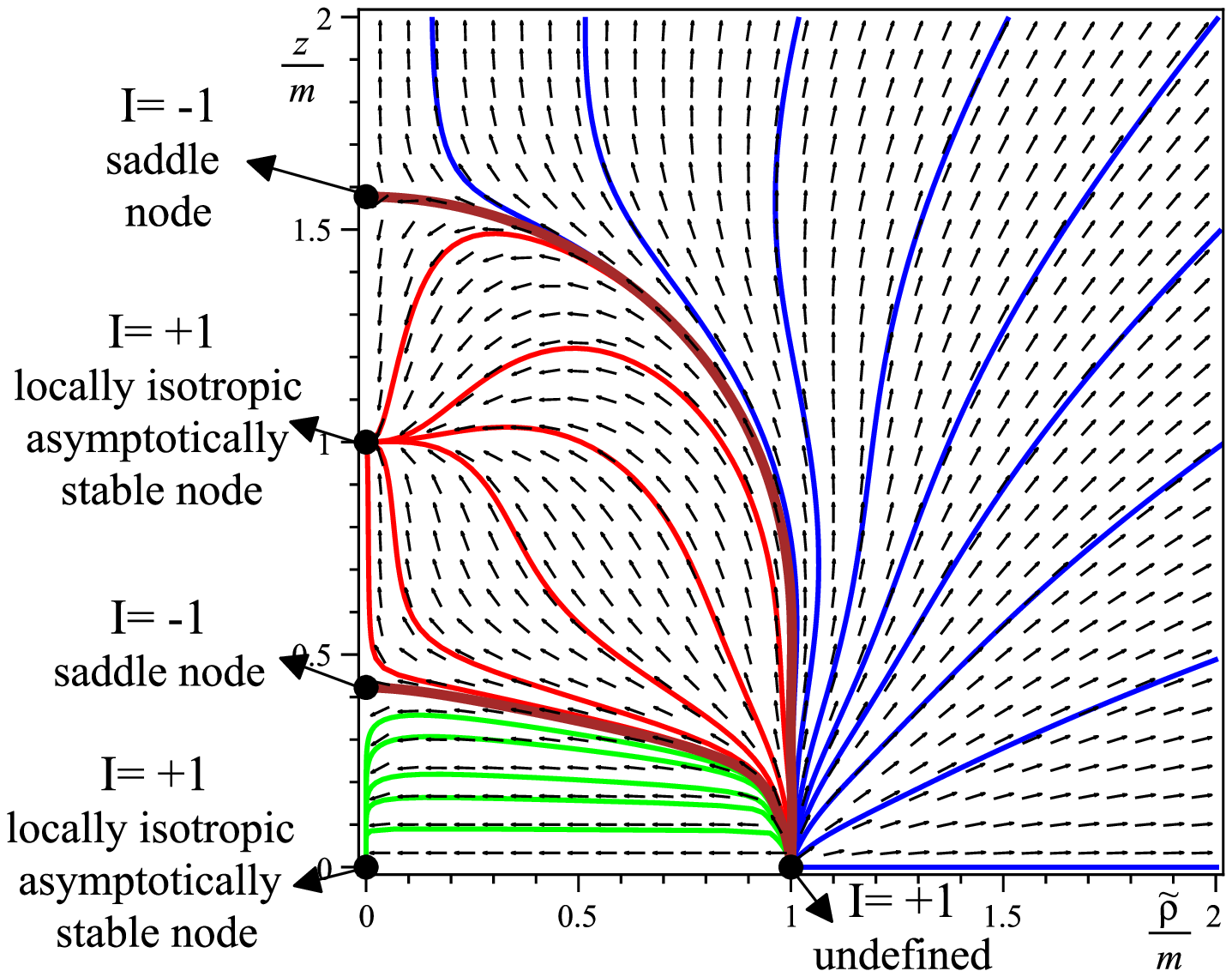,height=3in,width=4in,angle=0}
\caption{\label{curzonnewA}As in Figure \ref{curzonA} but with the new unfolding. }
\end{figure}

\begin{figure}[ht]
\epsfig{file=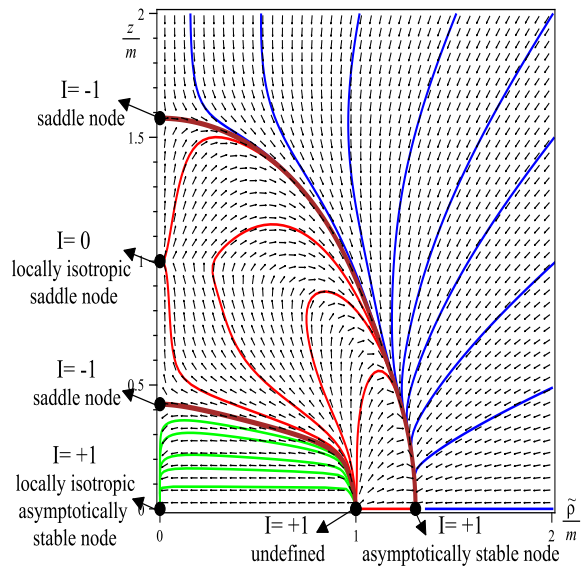,height=3in,width=4in,angle=0}
\caption{\label{curzonnewB}As in Figure \ref{curzonB} but with the new unfolding.}
\end{figure}

Whereas the unfolding of $(\rho,z)=(0,0)$ given by Scott and Szekeres accomplishes the task, the rather complicated procedure also modifies the entire spacetime representation. Here we seek a new unfolding of $(\rho,z)=(0,0)$ which does not modify the spacetime in the large. It turns out that we need only modify $\rho$. On reviewing Appendix A we see that the most important term to consider is the exponent of $\sin(\theta)/x$ (where $x \equiv r/m$). We can write
\begin{equation}\label{exponent}
    \frac{\sin(\theta)}{x}=\frac{\rho m}{r^2}.
\end{equation}
Now either this exponent diverges or it does not. If it diverges we can compactify this divergence with a $\tanh$ function. If it does not diverge we can set the term to zero by multiplying by $\rho$. Finally, let us require that the new $\rho$ and old $\rho$ approach each other for sufficiently large $r$. We arrive at the unfolding
\begin{equation}\label{transform}
    \frac{\tilde{\rho}}{m} = \frac{\rho}{m} + \tanh\left( \frac{\rho}{m} e^{\rho m/r^2}\right)\left(1-\tanh(\frac{r}{m})\right).
\end{equation}
The singularity is at $(\tilde{\rho},0) = (m,0)$ and the critical point is at $(\tilde{\rho},0) = (0,0)$. No flow lines cross the ``edge" $0 < \tilde{ \rho} < m$ where $r=0$. Indeed, whereas the trajectories $\theta=0$ and $\theta= \pi$ reach $\tilde{\rho} = 0$, all other trajectories of constant $\theta$ reach $\tilde{\rho} = m$. This is shown in figure \ref{newcoords}. The gradient fields of the CC metric are shown in Figures \ref{curzonnewA} and \ref{curzonnewB} with this new unfolding. It is very important to realize that this unfolding cannot correct all misrepresentations created by the Weyl coordinates. In particular, if the Weyl coordinates do not cover the ``edge", neither does the unfolding.

\subsection{``mass"}
Whereas the constant $m$ has entered the CC solution via the Newtonian potential (\ref{ccu}), the meaning of $m$ within the CC solution is no longer obvious. This is explored in Appendix D where we show that $m$ is certainly the ``mass" at spatial infinity. However, away from spatial infinity, looking at quasi local and local constructions, we find that the Hawking mass $\mathcal{M}_{H}$ provides no useful information for the CC solution for small $r$. Rather, it is the classical effective gravitational mass $\mathcal{M} \equiv \mathcal{R}_{\theta \;\phi}^{\;\;\;\;\; \theta \; \phi}g_{\theta \theta}^{\;\;\;3/2}/2$ \cite{lake1} that provides useful information on the CC solution. Whereas this $\mathcal{M}$ rapidly converges to $m$ with increasing $r$, near $r=0$, $\mathcal{M}$ shows considerable structure. Most interesting is the fact that $\mathcal{M}=0$ at the (naked) singularity, reminiscent of spherically symmetric naked singularities \cite{lake2}.

\section{The Newtonian analogue}
\subsection{Construction}
As explained previously \cite{lake1}, we construct the Newtonian tidal tensor
\begin{equation}\label{tide}
    E_{a b}=\Phi_{,a ,b}-\frac{1}{3}\eta_{a b} \square \;\Phi,
\end{equation}
and define the associated invariant
\begin{equation}\label{newinvar}
   \mathcal{I}_{1} = E_{a b}E^{a b}=\Phi_{,a ,b} \Phi^{,a ,b} -\frac{1}{3}(\square \;\Phi)^2.
\end{equation}
We now construct the gradient field
\begin{equation}\label{fields}
    l_{1\; c} \equiv - \nabla_{c}(E_{a b}E^{a b}).
\end{equation}
We say that $l_{1}$ is a Newtonian analogue (in no way any limit) of $k_{1}$ if their associated phase portraits are ``analogous", a designation which is explained quantitatively in detail below. This analogy is strict since, as explained in \cite{lake1}, $k_{1 \;\alpha} = - \nabla_{\alpha}(E_{\beta \gamma}E^{\beta \gamma})$ where $E_{\beta \gamma}$ is the usual electric component of the Weyl tensor; that is, the general - relativistic tidal tensor in the Ricci - flat case. The generalization of (\ref{fields}) for comparison with $k_{2}$ is
\begin{equation}\label{fields2}
    l_{2\; d} \equiv - \nabla_{d}(E_{a}^{b}E_{b}^{c}E_{c}^{a}),
\end{equation}
but, we caution, the physical meaning of the associated scalar is not known.

The Newtonian potential for a infinitely thin ring in vacuum with radius $a$ and mass $m$ is
\begin{equation}
\Phi_{ring}(\rho,z) = -\frac{\tilde{m}}{2\pi}\int_0^{2\pi}{\frac{d\theta}{\sqrt{{\tilde{\rho}}^2+{\tilde{z}}^2+1-2\tilde{\rho}\cos{\theta}\;}\;}},
\end{equation}
where $\tilde{x} \equiv x/a \;\; \forall \;x$. Note that $m$ only affects the intensity of the gradient field, but does not change the shape of the flow lines or the normalized field when the coordinates are parameterized by the radius $a$. We now drop $\tilde{}$ and take $m=1$.

\begin{figure}[ht]
\epsfig{file=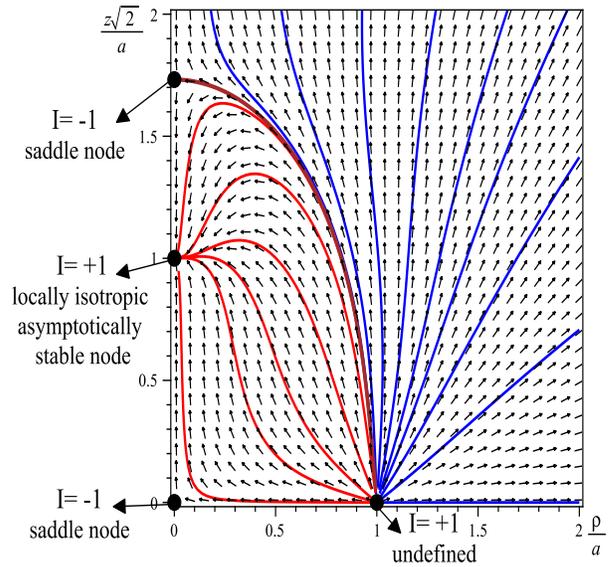,height=3in,width=4in,angle=0}
\caption{\label{ringA}Gradient field $l_{1\; a}$, defined by (\ref{fields}), for a Newtonian ring in vacuum. We have used $\sqrt{2}z$ for visualization purposes only. Compare Figure \ref{curzonnewA}.}
\end{figure}

\begin{figure}[ht]
\epsfig{file=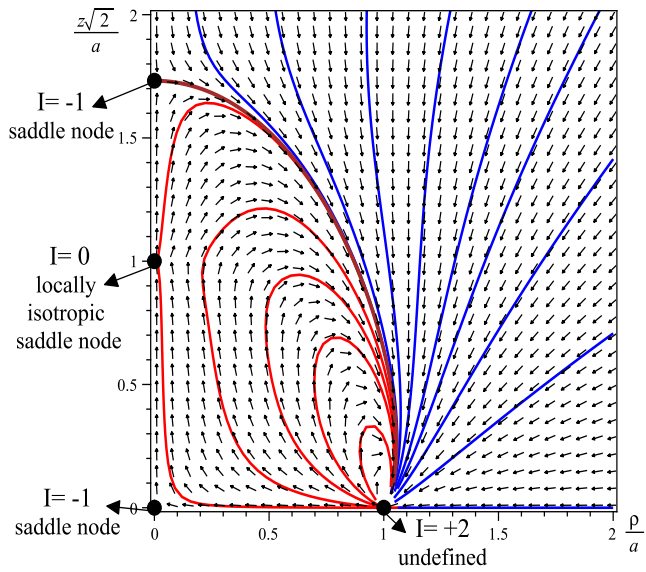,height=3in,width=4in,angle=0}
\caption{\label{ringB}Gradient field $l_{2\; a}$, defined by (\ref{fields2}), for a Newtonian ring in vacuum. Compare Figure \ref{curzonnewB}.}
\end{figure}

Even though we are in Newtonian vacuum ($\square \;\Phi=0$), the formulae off $\rho =0$ are too large to give here (we give the general formulae in Appendix E). Along $\rho=0$ we find
\begin{equation}\label{newtk}
    l_{1\; a}=\frac{7z(2z^2-1)(2z^2-3)}{(z^2+1)^6} \delta_{a}^z,
\end{equation}
and so we have critical points at $z = 0, \pm 1/\sqrt{2}$ and $\pm\sqrt{3/2}$. The non -zero components of the Hessian are given by
\begin{equation}\label{hessn1}
    H_{\rho \rho} = \frac{-28z^6+116z^4-39z^2+12}{(z^2+1)^7}
\end{equation}
and
\begin{equation}\label{hessn2}
    H_{z z} = \frac{7(28z^6-92z^4+57z^2-3)}{(z^2+1)^7}
\end{equation}
so that the determinant is given by $H=H_{\rho \rho}H_{z z}$. Finally, again along $\rho = 0$, we find
\begin{equation}\label{In}
    \mathcal{I}_{1} = \frac{7(2z^2-1)^2}{6(z^2+1)^5}.
\end{equation}
Since $E_{a b}$ vanishes at the critical points $z= \pm 1/\sqrt{2}$, and borrowing the notation from the relativistic case, we call these critical points ``isotropic". Otherwise the critical points are classified in the usual way. The Newtonian gradient field, obtained numerically, is shown in Figure \ref{ringA}. The overall index is $+1$. Following an analogous procedure for $l_2$ we obtain Figure \ref{ringB}. Again the overall index is $+1$.

\subsection{Comparison}
Now in comparing Figures \ref{curzonnewA} and \ref{ringA} we need to be quantitative, not just qualitative. It is important to recall that the existence of a critical point is coordinate independent (up to the use of defective coordinates) as is the classification of critical points. We say that two flows are analogous, within respective regions, if the flows give the same number and type and order of critical points and, therefore, the same Euler characteristic for the regions. Once again, this comparison is coordinate independent. Comparing Figures \ref{curzonnewA} and \ref{ringA} we see that the flow in the CC metric, excluding the inner (green) region, is analogous to the flow associated with a Newtonian ring. An unfolding of the CC metric at $(\rho,z)=(0,0)$ is central to this comparison. The fact that the CC solution contains an additional flow (the inner (green) flow) not present for a Newtonian ring is a fact we simply have to accept. We must also accept the fact that the CC solution itself is incompletely understood in this region.

Comparing Figures \ref{curzonnewB} and \ref{ringB} we see that as well as the additional inner flow (green) in the CC metric, there is also an additional critical point in the equatorial plane not present for a Newtonian ring. Otherwise, as is evident, the gradient flows are remarkably similar. Since the physics of (\ref{fields2}) has not been established, we content ourselves with a comparison of Figures \ref{curzonnewA} and \ref{ringA}.

\section{Discussion and Conclusion}

We have demonstrated a new tool designed to visualize spacetime curvature based on the construction of gradient flows of invariants. The emphasis here has been on the construction of strict Newtonian analogues wherein the invariants are the relativistic and Newtonian tidal invariants. The case we have considered, the Curzon - Chazy solution, is by no means an easy case to start out with. Despite this, we have shown that the gradient flows for the tidal invariant in the Curzon - Chazy solution, and that for an infinitely thin Newtonian ring, are, in a quantitative sense, remarkably similar in detail. This quantitative comparison of regions involves the number, type and order of critical points (and so the associated Euler characteristic). This comparison is coordinate independent. Only ``interior to" and ``close to" the Curzon - Chazy ``ring" (within the framework of a new unfolding of the Curzon - Chazy solution)  do the flows differ. This region is absent in the Newtonian case. For completeness we have included all non - differential invariants of the Curzon - Chazy solution and so we have also included a study of the second Weyl invariant and its Newtonian counterpart for an infinitely thin ring. Again we find that the gradient flows are remarkably similar in detail except for the same region close to the Curzon - Chazy ring and the addition of an external, but ``close",  critical ``ring" of the flow in the equatorial plane not present for the Newtonian ring. The arguments presented here, we believe, go a long way to clarify the notion that the ``source" of the Curzon - Chazy solution is ``ring - like" and that the construction of strict Newtonian analogues, correct ``in the large", is possible.

\begin{acknowledgments}
This work was supported in part by a Duncan and Urlla Carmichael Graduate Fellowship (to MA) and a grant (to KL) from the Natural Sciences and Engineering Research Council of Canada. Portions of this work were
made possible by use of \textit{GRTensorII} \cite{grt}.
\end{acknowledgments}

\newpage

\appendix
\begin{widetext}
\section{Stachel Coordinates}
Under the transformations
\begin{equation}\label{polar}
    \rho = r \sin(\theta),\;\;\;z=r \cos(\theta)
\end{equation}
the CC metric takes the form

\begin{equation}\label{polarmetric}
    ds^2 = -e^{-\frac{2m}{r}}dt^2+e^{\frac{2m}{r}}(e^{-\frac{m^2 \sin(\theta)^2}{r^2}}(dr^2+r^2 d \theta^2)+r^2 \sin(\theta)^2 d \phi^2).
\end{equation}

Either from (\ref{polarmetric}), or from CC metric and (\ref{polar}), we obtain
\begin{equation}\label{w1rpolar}
    w1 \equiv (w1R) m^4=2e^{2t1}\frac{(3t2+\sin(\theta)^2t3)}{x^{10}}
\end{equation}
and
\begin{equation}\label{w2rpolar}
    w2 \equiv (w2R) m^6=-3e^{3t1}\frac{(x-1)(2t2+\sin(\theta)^2t3)}{x^{14}}
\end{equation}
where
\begin{equation}\label{x}
    x \equiv r/m,
\end{equation}

\begin{equation}\label{t1}
    t1 \equiv \frac{\sin(\theta)^2-2x}{x^2},
\end{equation}

\begin{equation}\label{t2}
    t2 \equiv x^2(x-1)^2,
\end{equation}
and
\begin{equation}\label{t3}
    t3 \equiv 3x^2-3x+1.
\end{equation}
Now let us take
\begin{equation}\label{adef}
    \sin(\theta) \equiv a(x),
\end{equation}
where $0 \leq a \leq 1$, in order to define trajectories $\theta(x)$. We are interested in limits as $x \rightarrow 0$. First let us assume $a \in C^2$.

\bigskip

\textit{i) $a(0)=a_{o} \neq 0$}

\bigskip

We find
\begin{equation}\label{w1lim}
    w1 \sim 2e^{2\frac{a_{o}^2}{x^2}}\frac{a_{o}^2}{x^{10}} \rightarrow \infty,
\end{equation}
and
\begin{equation}\label{w2lim}
    w2 \sim -3e^{3\frac{a_{o}^2}{x^2}}\frac{a_{o}^2}{x^{14}} \rightarrow -\infty.
\end{equation}

\textit{ii) $a(0)=0$}

\bigskip

We find
\begin{equation}\label{w1lim1}
    w1 \sim \frac{6}{e^{\frac{4}{x}}x^8} \rightarrow 0,
\end{equation}
and
\begin{equation}\label{w2lim1}
    w2 \sim -\frac{6}{e^{\frac{6}{x}}x^{12}} \rightarrow 0.
\end{equation}

Historically, $a \in C^0$ have played a role. Following \cite{Cooperstock}, in the notation of \cite{Scott1986}, consider the trajectories
\begin{equation}\label{special}
   \frac{z}{m}  = b(\frac{1}{2})^{\frac{1}{3}}(\frac{\rho}{m})^n, \;\;\; (b,n>0),
\end{equation}
that is,
\begin{equation}\label{special1}
    x \cos(\theta) = b(\frac{1}{2})^{\frac{1}{3}}(x \sin(\theta))^n.
\end{equation}
Now if $a(0)=0$, we can use the small angle formula to obtain
\begin{equation}\label{small}
    a = \frac{x^{\frac{1-n}{n}}}{B^{\frac{1}{n}}},\;\;\;B \equiv  b(\frac{1}{2})^{\frac{1}{3}}.
\end{equation}
Of particular interest is the case $n=2/3$ for which
\begin{equation}\label{crit}
    a=\left(\frac{2x}{b^3}\right)^{\frac{1}{2}}.
\end{equation}
For the convergence of $w1$ and $w2$ we must have
\begin{equation}\label{converg}
    \frac{a^2-2x}{x^2} < 0,
\end{equation}
that is, $b>1$ for the case (\ref{crit}). This is the correction in \cite{Scott1986} to an error in \cite{Cooperstock}.
Further details concerning the very particular choice (\ref{special}) can be found in \cite{Scott1986}. Here we simply note that $a \in C^1$ requires $n < 1/2$ and $a \in C^2$ requires $n < 1/3$ for this particular choice. In \cite{Scott1986}, Scott and Szekeres go on to argue (on page 562) that they found a trajectory along which $w1R$ goes to a finite non - zero constant as $r \rightarrow 0$. We have examined this claim in detail and find the claim to be false. Along the suggested trajectory we find that the scalar $w1$ goes to zero.\footnote{This is rather unfortunate in the sense that had the claim been correct, a further unfolding of the singularity would be absolutely necessary for a complete understanding of the singularity.} Our understanding is that only two limits are possible within known coordinates: zero and $\pm$ infinity.

\bigskip

Now for the metric (\ref{polarmetric}), $\xi^{\alpha}=\delta^{\alpha}_{t}$ is a Killing vector. As a result, for all geodesics with tangents $u^{\alpha}$ we have a constant of the motion $\xi^{\alpha}u_{\alpha} \equiv -\gamma$. As a result, for all geodesics we have
\begin{equation}\label{geodesics}
    \frac{d t}{d \lambda} = \pm \gamma e^{\frac{2}{x}}
\end{equation}
where $\lambda$ is an affine parameter. This shows the inadequacy of the coordinate $t$ as $x \rightarrow 0$.
\section{Gradient Fields in Stachel Coordinates}
The gradient flow associated with the invariant $w1r$ is given by
\begin{equation}\label{krs}
    m^5k_{1\;r}=2e^{2t1}\frac{\sin(\theta)^2(4 \sin(\theta)^2t3+t4)+6t5}{x^{13}}
\end{equation}
and
\begin{equation}\label{kthetas}
m^6k_{1\;\theta}=- 4e^{2t1}\frac{\sin(\theta) \cos(\theta)(2 \sin(\theta)^2t3+t6)}{x^{14}}
\end{equation}
where
\begin{equation}\label{t4}
    t4 \equiv x(36x^3-63x^2+34x-4),
\end{equation}
\begin{equation}\label{t5}
    t5 \equiv x^3(3x^2-6x+2)(x-1),
\end{equation}
and
\begin{equation}\label{t6}
    t6 \equiv x^2(9x^2-15x+7).
\end{equation}
The gradient flow associated with the invariant $w2r$ is given by

\begin{equation}\label{kr}
    m^7k_{2\;r}=- 3e^{3t1}\frac{6(x-1)t5+\sin(\theta)^2(6\sin(\theta)^2(x-1)t3+xt7)}{x^{17}}
\end{equation}
where
\begin{equation}\label{t7}
    t7 \equiv 45x^4-126x^3+124x^2-50x+6,
\end{equation}
and
\begin{equation}\label{ktheta}
m^8k_{2\;\theta}= 6e^{3t1}\frac{\sin(\theta) \cos(\theta)(x-1)(3 \sin(\theta)^2t3+t6)}{x^{18}}.
\end{equation}

\section{Scott - Szekeres Unfolding}
The unfolding of $(\rho,z)=(0,0)$ given by Scott and Szekeres in \cite{Scott1986} and \cite{Scott1986a}, obtained by trial and error, is

\begin{equation}\label{ssxw}
X= \arctan((\rho/m)\,e^{m/z})+\arctan((\rho/m)\,e^{(-\sqrt{2}m/\rho)^{2/3}})
\end{equation}
and
\begin{equation}\label{ssyw}
Y=\arctan\left(  3\frac{z}{m}-\frac{(z/m)^2\, e^{m/r-m^2 \rho^2/2r^4})}{((r/m)^8+1+\frac{1}{3}(\rho/m)^2\,(r/m)^{-4})^{1/4}} \right)
\end{equation}
in Weyl coordinates. In Stachel coordinates we find
\begin{equation}\label{ssxs}
   X=\arctan(x \sin(\theta)e^{1/x \cos(\theta)})+\arctan(x \sin(\theta)e^{(-\sqrt{2}/ x\sin(\theta))^{2/3}})
\end{equation}
and
\begin{equation}\label{ssys}
  Y= \arctan\left(3 x \cos(\theta)-\frac{x^2 \cos(\theta)^2e^{-t1/2}}{(x^8+1+\sin(\theta)^2/3x^2)^{1/4}}\right).
\end{equation}

\section{``mass"}
We are interested to see how the constant $m$ in the CC solution is related to ``mass". First, let us rewrite (\ref{polarmetric}) via Taylor series about $1/r=0$ with explicit terms to order $1/r$. We have
\begin{equation}\label{taylor}
    ds^2 = -(1-\frac{2m}{r})dt^2+(1+\frac{2m}{r})(dr^2+r^2d\Omega^2_2)
\end{equation}
where $d\Omega_{2}^2$ is the metric of a unit 2-sphere ($d\theta^2+\sin^2(\theta)d\phi^2$). To this order in $r$ we find that the Einstein tensor for (\ref{taylor}) vanishes. Due to the spherical symmetry of (\ref{taylor}) we can consider the mass defined by $\mathcal{M} \equiv \mathcal{R}_{\theta \;\phi}^{\;\;\;\;\; \theta \; \phi}g_{\theta \theta}^{\;\;\;3/2}/2$ \cite{lake1}. To order $1/r$ we find
\begin{equation}\label{taylormass}
    \frac{\mathcal{M}}{m} = 1-\frac{3}{2}\left(\frac{1}{x}\right).
\end{equation}
As a result, at spatial infinity, $m$ is the ``mass". We need not consider the ADM mass, the Komar integrals nor the Bondi -Sachs mass \cite{poisson}. Rather, we are interested in exploring the meaning of $m$ away from spatial infinity. We are, therefore, interested in local and quasi-local quantities.

Let us look at the Hawking mass \cite{Szabados}, which can be defined by
\begin{equation}\label{hawking}
    \mathcal{M}_{H} = \left(\frac{A(\mathcal{S})}{16 \pi}\right)^{1/2}\left(1-\frac{1}{2 \pi}\oint_{\mathcal{S}}\rho \mu d \mathcal{S}\right)
\end{equation}
where $\mathcal{S}$ is a spacelike topological two-sphere, $A(\mathcal{S})$ is the associated area and here $\rho$ and $\mu$ are the Newman - Penrose spin coefficients. As mentioned above, Stachel \cite{Stachel1968} showed that for $\mathcal{S}$ defined by surfaces of constant $t$ and $r$ in (\ref{polarmetric}), $A$ decreases with decreasing $r$ up to a minimum (which we find to be $r/m \simeq 0.5389$) and then diverges as $r \rightarrow 0$. To ensure that $\mathcal{S}$ is a  topological two - sphere we use the standard Gauss - Bonnet theorem
\begin{equation}\label{gauss}
    \frac{1}{2}\int \int _{\mathcal{S}} \mathcal{R} \sqrt{g} dx^a dx^b = 2 \pi \chi(\mathcal{S})
\end{equation}
where $x^a$ are the coordinates on $\mathcal{S}$, $g$ is the determinant of the metric on $\mathcal{S}$, $\mathcal{R}$ is the Ricci scalar on $\mathcal{S}$, and $\chi(\mathcal{S})$ is the Euler characteristic of $\mathcal{S}$. Again taking $\mathcal{S}$ to be defined by surfaces of constant $t$ and $r$ in (\ref{polarmetric}) we find
\begin{equation}\label{euler}
    \chi(\mathcal{S})=2,
\end{equation}
and so we are indeed considering topological two - spheres.
Constructing a complex null tetrad in the usual way, we find that for (\ref{polarmetric})
\begin{equation}\label{rhmu}
    -\rho \mu = -\frac{1}{8}\frac{\exp{(\frac{m^2 \sin(\theta)^2}{r^2})}(-2mr+2r^2+m^2 \sin(\theta)^2)^2}{e^{(\frac{m}{r})^2}r^6}
\end{equation}
and so we find
\begin{equation}\label{hawkingcurzon}
\frac{\mathcal{M}_{H}}{m}=-1/32\,\sqrt {2}{e^{y}}\sqrt {\int _{0}^{\pi }\!{\frac {\sin \left(
\theta \right) }{\sqrt {{e^{ \left( \sin \left( \theta \right)
 \right) ^{2}{y}^{2}}}}}}{d\theta}} \left( -8-2\,\sqrt {2}{y}^{2}\pi -
3/4\,\sqrt {2}{y}^{4}\pi +{\frac {5}{32}}\,\sqrt {2}{y}^{5}\pi +7/4\,
\sqrt {2}{y}^{3}\pi +\sqrt {2}y\pi  \right) {y}^{-1}
\end{equation}
where $y \equiv m/r =1/x$. Expanding about $y=0$ we find that
\begin{equation}\label{hawking expansion}
\frac{\mathcal{M}_{H}}{m}=1-\frac{1}{15x^{3}}-\frac{2}{315x^5}+\mathcal{O}(\frac{1}{x^6}).
\end{equation}
Examining (\ref{hawkingcurzon}) in more detail we find that $\mathcal{M}_{H}=0$ for $r/m \sim .5742$ and $\mathcal{M}_{H} \rightarrow -\infty$ as $r \rightarrow 0$.  In this regard, one is reminded of the result of Hansevi \cite{hansevi} who showed that the Hawking mass can be negative even for convex two - surfaces in Minkowski spacetime. We have to conclude that $\mathcal{M}_{H}$ is not a good measure of ``mass" for the CC solution except for large $r$ (in fact $\mathcal{M}_{H}$ converges very rapidly to $m$, for example, $\mathcal{M}_{H}/m  \simeq .9914$ at $x=2$).

Returning to $\mathcal{M} \equiv \mathcal{R}_{\theta \;\phi}^{\;\;\;\;\; \theta \; \phi}g_{\theta \theta}^{\;\;\;3/2}/2$, whereas this quantity is usually restricted to strict spherical symmetry, it is well defined for (\ref{polarmetric}). Indeed, a straightforward calculation gives
\begin{equation}\label{regularmass}
    \frac{\mathcal{M}}{m} = e^{-\frac{t1}{2}}\left(1+\frac{t1}{2}\right),
\end{equation}
where $t1$ is given by (\ref{t1}). Continuing with (\ref{adef}) we find
\begin{equation}\label{m1lim}
    \lim_{x \rightarrow 0}\left( \frac{\mathcal{M}}{m}\right) \rightarrow 0,
\end{equation}
for $a(0)=a_{o} \neq 0$ and
\begin{equation}\label{m2lim}
    \lim_{x \rightarrow 0}\left( \frac{\mathcal{M}}{m}\right) \rightarrow -\infty,
\end{equation}
for $a(0)=a_{o} = 0$. In strict spherical symmetry, it is known that naked singularities are massless ($\mathcal{M}=0$) \cite{lake2}. The result (\ref{m1lim}) suggests that this might hold away from spherical symmetry. It should be noted that $\frac{\mathcal{M}}{m}$ converges very rapidly to $1$ for all $\theta$ with increasing $x$. We conclude that $\mathcal{M}$, if not the ``mass", at least summarizes some important properties of the CC solution.
\section{Newtonian $\mathcal{I}_{1}$ for vacuum}
In Newtonian vacuum we have $\square \;\Phi(\rho,z)=0$. The components of $E_{a b}$ are obvious. For $\mathcal{I}_{1}$ we find
\begin{equation}\label{newtoni1}
\mathcal{I}_1 = {\frac {( {\frac {\partial ^{2}}{\partial {\rho}^{2}}}\Phi
 ) ^{2}{\rho}^{2}+2\, ( {\frac {
\partial ^{2}}{\partial \rho\partial z}}\Phi
 ) ^{2}{\rho}^{2}+ ( {\frac {\partial ^{2}}{\partial {z}^{2
}}}\Phi  ) ^{2}{\rho}^{2}+ ( {\frac {
\partial }{\partial \rho}}\Phi ) ^{2}}{{
\rho}^{2}}}.
\end{equation}
The components of $l_{1\; a}$ are
\begin{eqnarray}\label{l1rho}
l_{1\; \rho} = -\frac{2}{\rho^3}\, (( {\frac {\partial ^{2}}{\partial {\rho}^
{2}}}\Phi ) {\rho}^{3}{\frac {\partial ^{
3}}{\partial {\rho}^{3}}}\Phi  +2\, ( {
\frac {\partial ^{2}}{\partial \rho\partial z}}\Phi  ) {\rho}^{3}{\frac {\partial ^{3}}{\partial {\rho}^{2}
\partial z}}\Phi  +  ( {\frac {\partial ^{2}}
{\partial {z}^{2}}}\Phi ){\rho}^{3}{
\frac {\partial ^{3}}{\partial z\partial \rho\partial z}}\Phi +\rho\, ( {\frac {\partial }{\partial \rho}}\Phi
 ) {\frac {\partial ^{2}}{\partial {\rho}
^{2}}}\Phi  - ( {\frac {\partial }{\partial
\rho}}\Phi ) ^{2}),
\end{eqnarray}
and
\begin{eqnarray}\label{l1z}
l_{1\; z} = -\frac{2}{\rho^2}\, (( {\frac {\partial ^{2}}{\partial {\rho}^{2}}}\Phi ) {\rho}^{2}{\frac {\partial ^{3}}
{\partial {\rho}^{2}\partial z}}\Phi +2\,
 ( {\frac {\partial ^{2}}{\partial \rho\partial z}}\Phi ) {\rho}^{2}{\frac {\partial ^{3}}{\partial z
\partial \rho\partial z}}\Phi +  ( {\frac {
\partial ^{2}}{\partial {z}^{2}}}\Phi ) {
\rho}^{2}{\frac {\partial ^{3}}{\partial {z}^{3}}}\Phi + ( {\frac {\partial }{\partial \rho}}\Phi ) {\frac {\partial ^{2}}{\partial \rho\partial z}}\Phi ).
\end{eqnarray}
The components of the Hessian are
\begin{eqnarray}
H_{\rho \rho} = \frac{2}{\rho^4}(\, ( {\frac {\partial ^{3}}{\partial {\rho}^{3}}}\Phi
   ) ^{2}{\rho}^{4}+ ( {\frac {
\partial ^{2}}{\partial {\rho}^{2}}}\Phi
 ) {\rho}^{4}{\frac {\partial ^{4}}{\partial {\rho}^{4}}}\Phi
  +2\, ( {\frac {\partial ^{3}}{\partial {
\rho}^{2}\partial z}}\Phi  ) ^{2}{\rho}^{4
}+2\,( {\frac {\partial ^{2}}{\partial \rho\partial z}}\Phi
  ) {\rho}^{4}{\frac {\partial ^{4}}{
\partial {\rho}^{3}\partial z}}\Phi  +   ( {
 \frac {\partial ^{3}}{\partial z\partial \rho\partial z}}\Phi  ) ^{2}{\rho}^{4}+ \\ \nonumber ( {\frac {\partial ^{2}}{
\partial {z}^{2}}}\Phi   ) {\rho}^{4}{
\frac {\partial ^{4}}{\partial z\partial {\rho}^{2}\partial z}}\Phi
  -4\,\rho\, ( {\frac {\partial }{\partial
\rho}}\Phi   ) {\frac {\partial ^{2}}{
\partial {\rho}^{2}}}\Phi  + ( {\frac {
\partial ^{2}}{\partial {\rho}^{2}}}\Phi
 ) ^{2}{\rho}^{2}+{\rho}^{2} ( {\frac {\partial }{\partial
\rho}}\Phi   ) {\frac {\partial ^{3}}{
\partial {\rho}^{3}}}\Phi  +3\, ( {\frac {
\partial }{\partial \rho}}\Phi  ) ^{2}),
\end{eqnarray}
\begin{eqnarray}
H_{\rho z} =\frac{2}{\rho^3}(( {\frac {\partial ^{3}}{\partial {\rho}^{2}\partial
z}}\Phi  ) {\rho}^{3}{\frac {\partial ^{3}
}{\partial {\rho}^{3}}}\Phi  + ( {\frac {
\partial ^{2}}{\partial {\rho}^{2}}}\Phi
) {\rho}^{3}{\frac {\partial ^{4}}{\partial {\rho}^{3}\partial
z}}\Phi +2\, ( {\frac {\partial ^{3}}{
\partial z\partial \rho\partial z}}\Phi
 ) {\rho}^{3}{\frac {\partial ^{3}}{\partial {\rho}^{2}\partial
z}}\Phi  +2\, ( {\frac {\partial ^{2}}{
\partial \rho\partial z}}\Phi ) {\rho}^{3
}{\frac {\partial ^{4}}{\partial z\partial {\rho}^{2}\partial z}}\Phi
 + \\ \nonumber ( {\frac {\partial ^{3}}{\partial {z}^{3
}}}\Phi   ) {\rho}^{3}{\frac {\partial ^{3}
}{\partial z\partial \rho\partial z}}\Phi  +
 ( {\frac {\partial ^{2}}{\partial {z}^{2}}}\Phi ) {\rho}^{3}{\frac {\partial ^{4}}{\partial {z}^{2}
\partial \rho\partial z}}\Phi +\rho\, ( {
\frac {\partial ^{2}}{\partial \rho\partial z}}\Phi  ) {\frac {\partial ^{2}}{\partial {\rho}^{2}}}\Phi
 +\rho\, ( {\frac {\partial }{\partial \rho
}}\Phi ) {\frac {\partial ^{3}}{\partial
{\rho}^{2}\partial z}}\Phi  -2\, ( {\frac {
\partial }{\partial \rho}}\Phi ) {\frac {
\partial ^{2}}{\partial \rho\partial z}}\Phi),
\end{eqnarray}
\begin{eqnarray}
H_{z z} =\frac{2}{\rho^2}( ( {\frac {\partial ^{3}}{\partial {\rho}^{2}\partial
z}}\Phi   ) ^{2}{\rho}^{2}+ ( {\frac {
\partial ^{2}}{\partial {\rho}^{2}}}\Phi
 ) {\rho}^{2}{\frac {\partial ^{4}}{\partial z\partial {\rho}^{2
}\partial z}}\Phi  +2\, ( {\frac {\partial ^
{3}}{\partial z\partial \rho\partial z}}\Phi
) ^{2}{\rho}^{2}+  2\,( {\frac {\partial ^{2}}{\partial
\rho\partial z}}\Phi ) {\rho}^{2}{\frac {
\partial ^{4}}{\partial {z}^{2}\partial \rho\partial z}}\Phi  + \\ \nonumber ( {\frac {\partial ^{3}}{\partial {z}^{3}}}\Phi
) ^{2}{\rho}^{2}+ ( {\frac {
\partial ^{2}}{\partial {z}^{2}}}\Phi) {
\rho}^{2}{\frac {\partial ^{4}}{\partial {z}^{4}}}\Phi + ( {\frac {\partial ^{2}}{\partial \rho\partial z}}\Phi
 ) ^{2}+ ( {\frac {\partial }{
\partial \rho}}\Phi) {\frac {\partial ^{
3}}{\partial z\partial \rho\partial z}}\Phi)
\end{eqnarray}
and
\begin{eqnarray}
H_{\theta \theta} =2\,{\frac { ( {\frac {\partial ^{2}}{\partial {\rho}^{2}}}\Phi
  ) {\rho}^{3}{\frac {\partial ^{3}}{
\partial {\rho}^{3}}}\Phi  +2\, ( {\frac {
\partial ^{2}}{\partial \rho\partial z}}\Phi
 ) {\rho}^{3}{\frac {\partial ^{3}}{\partial {\rho}^{2}\partial
z}}\Phi  + ( {\frac {\partial ^{2}}{
\partial {z}^{2}}}\Phi ) {\rho}^{3}{
\frac {\partial ^{3}}{\partial z\partial \rho\partial z}}\Phi  +\rho\,( {\frac {\partial }{\partial \rho}}\Phi
  ) {\frac {\partial ^{2}}{\partial {\rho}
^{2}}}\Phi  -  {\frac {\partial }{\partial
\rho}}\Phi  ) ^{2}}{{\rho}^{2}}}.
\end{eqnarray}
\end{widetext}

\end{document}